\documentclass{article}
\usepackage{amsmath}
\usepackage{color}
\usepackage{amssymb}

\title{Interpretation of quantal manifolds}
\author{Johan Noldus}

\begin{document}

\maketitle

\begin{abstract}
In quantum gravity, one looks for alternative structures to spacetime
physics than ordinary real manifolds.  Here, we propose an
alternative universal construction containing the latter as an equilibrium
state under the action of the universal diffeomorphism group. Our theory
contains many other previous proposals in the literature as special cases.
However, the crucial point we make is that those have to be appreciated
in the universal context developed here.
\end{abstract}

\section{Introduction}
In modern physics, people question the very fabric of spacetime from many
different vantage points of view. As explained in great detail in the upcoming book of the author \cite{Noldus1}, the superposition principle cannot be applied to
spacetime which implies that spacetime cannot be quantized in the operational
sense. This indicates that an observer lives in one spacetime and since no a
priori discreteness can be imposed, the author \cite{Noldus2} reached the conclusion that
any approach to spacetime had to be based on the continuum given that the
notion of locality is only canonically defined in that context. This appears to
imply that even in the ontological sense, a standard real manifold is the only
natural candidate for a spacetime structure. So, the only question is whether
there exists a universal construction based on the continuum allowing for more
generic possibilities? The answer to this question surprisingly is yes and the
difference lies in the statistical density matrix approach to quantum mechanics
and the normal textbook state approach. We can, and will look at spacetime
in the quantum statistical sense with the standard notions of locality inherited
from $\mathbb{R}^4$ . The approach we will take is the algebraic one by means of $W^{\star}$ alge-
bras; this is just a temporary step and we are aware that more exotic avenues
will have to be taken as explained in \cite{Noldus1}. The problematic aspect of all non-
commutative approaches so far is that the diffeomorphism group has no natural
place in the formalism and indeed, imposing algebraic relations by hand breaks
diffeomorphism invariance of the single algebra. The answer to this problem
is to consider all possible algebras and modeling one manifold on a particular
one. Hence, a diffeomorphism will map one manifold into another and the only
fixed manifolds are the abelian and free ones. Moreover, the abelian continuum
spacetimes have the largest symmetry group and therefore they are preferred
from the point of view of internal symmetries. Therefore, any quantum space-
time dynamics should be based upon the fact that a maximal internal symmetry
group, as a subgroup of the free diffeomorphisms, determines the only stable
ground state. Hence, we conjecture that the theory developed in \cite{Noldus1} describes
the ground state of a much larger one which allows for small scale granularity
as quantum fluctuations at sufficiently small scales.
This paper is organized as follows: first we introduce topological manifolds and
give some examples. However, a deeper understanding of topological manifolds
emerges from the development of (first order) differential calculus; this is accomplished in section 3 and some nontrivial insights are provided. The construction
of higher differential operators, the curvature of the quantum connection and
the general definition of differentiable manifolds is postponed to future work.
Although this approach to quantum spacetime has been developed independently, the most valuable personal contact in this regard has been with Shahn
Majid, some of whose writings and ideas regarding $C^{\star}$ algebraic representations
of Hopf spacetime algebras have been useful. In particular, I recommend \cite{Majid1}
and \cite{Majid2}. Also, in retrospect, some of the ideas in this paper resemble those of
Grothendieck topology in the sense that the open algebras form a sheaf over
an ordinary topological space and the immersions of the open algebras associated to the open subsets in the covering of charts are prime examples of what
category theorists call a sieve. However, there is also more information to it
which is given in the definition of the local algebras attached to the generators
of the coordinate structure \cite{Kleinert}. Therefore, our construction is more restricted
than the one of Grothendieck topology (since we can also do analysis) and it
might be helpful to see if there exists a more category theoretical definition for
which our manifolds constitute particular representations. There is a very slight
resemblance to the standard Haag Kleinert axioms of Quantum Field Theory
which also works with local algebras over $\mathbb{R}^4$ but the correspondence does not
carry very far in the sense that no Minkowski causality or anything like that is
implemented.
\section{Topological quantum manifolds}
Basically, the universal complex (or real) algebra in $n$ variables $\widehat{x}_i$ is the free one $\mathcal{F}^{\infty}_n$; we shall 
also be concerned with the free algebra of finite words $\mathcal{F}_n$ which is equipped with a canonical involution $\star$ which 
simply reverses the order of the words and conjugates the complex numbers.  Hence, every generator is Hermitian and therefore has a 
real spectrum if one restricts to $W^{\star}$ algebraic representations.  Besides $\mathcal{F}^{r}_n$, there is the totally commutative 
algebra $\mathcal{C}^{r}_n$ in $n$ variables $x_i$ and we
 denote by $\phi : \mathcal{F}^{r}_n \rightarrow \mathcal{C}^{r}_n : \widehat{x}_i \rightarrow x_i$ the canonical homomorphisms 
where $r \in \{ \emptyset , \infty \}$.  Morever, we adjoin all algebras with an identity element and restrict to unital
 $\star$ homomorphisms.  The idea is to represent $\mathcal{F}_n$ in unital $W^{\star}$ algebras $\mathcal{A}$ equipped with a 
trace functional $\omega_{\mathcal{A}}$.  Therefore 
let $\pi : \mathcal{F}_n \subset \textrm{Dom}(\pi) \subset \mathcal{F}^{\infty}_n \rightarrow \mathcal{A}$ be a unital, maximal, 
star homomorphism (where $\textrm{Dom}(\pi)$ is a subalgebra) with a dense image and denote by $\sigma(i, \pi, \mathcal{A})$ the spectrum
 of $\pi(\widehat{x}_i)$ in $\mathcal{A}$; then it is natural to construct the compact and bounded ``cube''
$$
\mathcal{O}(\pi,\mathcal{A}) = \times_{i = 1}^{n} \sigma(i, \pi, \mathcal{A})\;.
$$
Likewise, one can restrict the variables in $\mathcal{C}_n$ to $\mathcal{O}(\pi, \mathcal{A})$.  Because of the spectral decompositon theorem, for every $n$ vector $\vec{\alpha}$ in the cube, index $i$ and $\epsilon_i > 0$, one has a unique Hermitian spectral operator $P^{\epsilon_i}_{\alpha_i}$ which is by definition a shorthand for 
$$
P^i (( \alpha_i - \epsilon_i, \alpha_i + \epsilon_i ))\;.
$$
The operators have the usual intersection properties.  Hence for every resolution $\vec{\epsilon}$, we may define an event $P^{\vec{\epsilon}}(\vec{\alpha})$ in the algebra $\mathcal{A}$ as the maximal Hermitian projection operator wich is smaller than all $P^{\epsilon_i}_{\alpha_i}$ (notice that this projection operator may become zero if the resolution becomes to high, that is $\epsilon_i$ too small).  Now, it is easy to see that if one were to cover a cube by smaller cubes (arbitrary overlaps are allowed), take the projection operators associated to those and consider the smallest projection operator which majorizes all of these, then, by the superposition principle, the latter is smaller or equal to the projection operator of the full cube.  This is a very quantum mechanical idea where we acknowledge that the whole is more than the sum of its parts and therefore we have to give up the idea of a classical partition.  Hence, for any relative open subset $\mathcal{W}$; there exists a unique smallest projection operator which majorizes all projection operators attached to subcoverings of $\mathcal{W}$ by relative open cubes (a subcovering simply is a set of relative open cubes contained in $\mathcal{W}$).  Hence, there is a natural almost everywhere weakly continuous\footnote{We shall explain this notion later on.} mapping $\kappa_{(\pi,\mathcal{A})}$ from relative open subets $\mathcal{W}$ of $\mathcal{O}(\pi,\mathcal{A})$ to $\mathcal{A}$ given by
$$
\kappa_{(\pi,\mathcal{A})}(\mathcal{W}) = P(\mathcal{W})\;.
$$
For disjoint $\mathcal{W}_j$ one obtains that
$$
P(\mathcal{W}_1)\, P(\mathcal{W}_2) = 0\;,
$$
meaning that the coherence of the theory depends upon the scale you are observing at.  Concretely, if you zoom into the region
 $\mathcal{W}_1$ you will be oblivious to the entanglement with the region $\mathcal{W}_2$; however, looking at both together 
gives a very different picture.  If the \emph{dynamics} itself were scale dependent in this way, then it might explain why we
 see a local world on our scales of observation and above, while the microscopic world would seem to be completely entangled.  This picture would offer a complete relativization of physics where giants would look to us as if we were electrons.  Also, 
$$
P(\mathcal{W}_1) \prec P(\mathcal{W}_2)
$$
of $\mathcal{W}_1 \subset \mathcal{W}_2$, which means that zooming in is a consistent procedure.  Now, we can go on and construct several forms of equivalence, going from ultra strong to ultra weak.  Two representations $\pi_i: \mathcal{F}_n \subset \textrm{Dom}(\pi_i) \subset \mathcal{F}^{\infty}_n \rightarrow \mathcal{A}_i$ are ultra strongly isomorphic if and only if there exists a $W^{\star}$ isomorphism $\gamma : \mathcal{A}_1 \rightarrow \mathcal{A}_2$ such that $\pi_2 = \gamma \circ \pi_1$ and $\textrm{Dom}(\pi_1) = \textrm{Dom}(\pi_2)$.  They are called strongly isomorphic it is only demanded that $\gamma$ is a unital star isomorphism from $\pi_1(\textrm{Dom}(\pi_1))$ to $\pi_2(\textrm{Dom}(\pi_2))$.  We say, moreover, that they are weakly isomorphic when equality is supposed to only hold on $\textrm{Dom}(\pi_1) \cap \textrm{Dom}(\pi_2)$ and finally we define them to be ultra weakly equivalent if and only if $\gamma$ is a star isomorphism from $\pi_1(\mathcal{F}_n)$ to $\pi_2(\mathcal{F}_n)$ and equality only holds on $\mathcal{F}_n$.  In the case of real manifolds, ultra weak equivalence is the only notion which applies and we continue now to investigate it.  Now, we are ready to go over to an atlas construction; a topological space $\mathcal{M}$ is said to be a real, $n$-dimensional, non-commutative manifold if there exists a covering of $\mathcal{M}$ by open sets $\mathcal{V}_{\beta}$, a homeomorphism $\phi_{\beta}$ from $\mathcal{V}_{\beta}$ to a relative open subset of the cube $\mathcal{O}(\pi_{\beta},\mathcal{A}_{\beta})$ associated to some representation $\pi_{\beta}: \textrm{Dom}(\pi_{\beta}) \rightarrow \mathcal{A}_{\beta}$ of the free algebra in n letters.  This homeomorphism canonically lifts to the algebra on the open subsets $\mathcal{W} \subset \mathcal{V}_{\beta}$ by stating that $\widehat{\phi}_{\beta}(\mathcal{W}) = \kappa_{(\pi_{\beta}, \mathcal{A}_{\beta})}(\phi_{\beta}(\mathcal{W}))$.  Hence, a single chart is a tuple $\left( \mathcal{V}_{\beta}, \pi_{\beta}, \mathcal{A}_{\beta}, \phi_{\beta} \right)$ and we proceed now to construct an atlas by demanding compatibility.  Two charts $\mathcal{V}_{\beta_j}$ with some non zero overlap $\mathcal{V}_{\beta_1} \cap \mathcal{V}_{\beta_2} \neq \emptyset$ are said to be compatible if and only if the canonical mapping between the normed subsets
$$
\{\widehat{\phi}_{\beta_{j}}(\mathcal{W}) \, | \, \mathcal{W} \subset \mathcal{V}_{\beta_1} \cap \mathcal{V}_{\beta_2} \}
$$
induces a star isomorphism between the normed algebras generated by them; the latter preserves the trace functionals $\omega_{\mathcal{A}_{\beta}}$.  We now proceed by giving some examples. \\* \\*
We start by the most trivial thing and show that ordinary real manifolds have a natural place in this setup.  Let $\mathcal{M}$ be
 an $n$-dimensional real manifold and consider the coordinate chart $(\mathcal{V}, \psi)$.  Define now the Hilbert 
space $L^2(\overline{\psi(\mathcal{V})}, d^n x)$ and the multiplication operators $x_i$.  Define $\mathcal{A}$ to be 
the $W^{*}$ algebra generated by the $x_i$, then $\pi : \mathcal{F}_n \rightarrow \mathcal{A} : \widehat{x}_i \rightarrow x_i$ 
has a unique maximal extension.  The spectrum of each of these multiplication operators is continuous and varies between $a_i < b_i$
 and the canonical mapping $\phi$ is given by $\phi(v) = \psi(v)$.  Then, the canonical projectors associated to
 $\mathcal{W} \subset \mathcal{V}$ are given by $\widehat{\phi}(\mathcal{W}) = \chi_{\phi (\mathcal{W})}$ where the latter is the characteristic 
function on $\phi(\mathcal{W})$.  Clearly, a coordinate tranformation induces a $W^{\star}$ algebraic isomorphism between these 
commutative projection operators.  By the same arguments, one sees that any commutative $n$-dimensional measure space is represented 
in this framework; so we are left with presenting a non abelian example.  A very simple example is a double sheeted manifold 
constructed from the Hilbert space L$^2(\mathbb{R}^4, d^4 x) \otimes \mathbb{C}^2$ and consider the algebra generated by the
 operators $x^{\mu} \otimes \sigma^{\mu}(x)$ where the $\sigma^{\mu}(x)$ are automorphic to the standard spacetime Pauli algebra
 $(\sigma^{\mu}) = (1, \sigma^i)$.  That is $\sigma^{\mu}(x) = U(x)\, \sigma^{\mu} U^{\dag}(x)$ for $U(x)$ some $2 \times 2$ complex 
unitary matrix.  The whole manifold structure depends upon $U(x)$, since suppose $U(x) = 1$, then the cube is
 $\mathbb{R}^4$ and the set of basic projection operators is given by
\begin{eqnarray*}
P^{\epsilon}_t & = & \chi_{\left[ t - \epsilon , t + \epsilon \right]} \otimes 1 \\
P^{\epsilon}_x & = & \frac{1}{2} \left[ \chi_{\left[ x - \epsilon, x + \epsilon \right]} \otimes |1,1 \rangle \langle 1,1 |  + \chi_{\left[ - x - \epsilon , - x + \epsilon \right]} \otimes | 1,-1 \rangle \langle 1, -1| \right]  \\
P^{\epsilon}_y & = & \frac{1}{2} \left[ \chi_{\left[ y - \epsilon, y + \epsilon \right]} \otimes |i,1 \rangle \langle i,1 |  + \chi_{\left[ - y - \epsilon , - y + \epsilon \right]} \otimes | -i,1 \rangle \langle -i, 1| \right] \\
P^{\epsilon}_z & = & \left[ \chi_{\left[ z - \epsilon, z + \epsilon \right]} \otimes |0,1 \rangle \langle 0,1 |  + \chi_{\left[ - z - \epsilon , - z + \epsilon \right]} \otimes | 1,0 \rangle \langle 1, 0| \right].   
\end{eqnarray*}
Hence, the operators $P^{\epsilon}_{(t,x,y,z)}$ vanish as soon as at least \emph{two} of the spatial coordinates have modulus greater
 or equal to $\epsilon$.  Therefore, if one is far away in two coordinates from the origin, one sees nothing except on the scales of 
the distances to the orgin itself.  If only one coordinate, say $z$ has a modulus greater than $\epsilon$, then the projection operator
 is given by 
\begin{eqnarray*}
P^{\epsilon}_{(t,x,y,z)} &=& \chi_{\left[ |x| - \epsilon, - |x| + \epsilon \right] \times \left[ |y| - \epsilon, - |y| + \epsilon \right] \times \left[ z - \epsilon, z + \epsilon \right]} \otimes |0,1 \rangle \langle 0,1 |  + \\
& & \chi_{\left[ |x| - \epsilon, - |x| + \epsilon \right] \times \left[ |y| - \epsilon, - |y| + \epsilon \right] \times \left[ - z - \epsilon , - z + \epsilon \right]} \otimes | 1,0 \rangle \langle 1, 0|\;,
\end{eqnarray*}
and the reader is invited to work out the projection operator for a case in which all spatial coordinates have a modulus smaller
 than $\epsilon$.   Therefore, one obtains an axial structure where any of the coordinate axes are priviliged which is logical since these are associated with maximal abelian subalgebra's. \\* \\*
In case $U(x) \neq 1$, the projection operators take on more complicated form due to different directions in spinor space.  Indeed, for $U(x) = e^{i\vec{a}(x).\vec{\sigma}}$ one has that $U(x)\sigma^i U(x)^{\dag}$ induces a rotation of an 
angle $2 ||\vec{a}|| \textrm{mod}
2 \pi$ around the vector $\vec{a}$; these can be computed exactly, as well as can the eigenvectors (although they are rather ugly 
functions of $\vec{a}$) and the latter are all of the type $v(x)\delta^n(y-x)$ where $v(x) \in \mathbb{C}^{2}$.  The reader may well 
have noticed that we still have to say something about dimension since dimensional collapse is possible; 
indeed any real $n$ dimensional manifold is a $m$ dimensional noncommutative one if and only if $m \geq n$.  On the other hand, 
discrete manifolds do not necessarily have a one dimensional representation due to the algebraic relations 
(so we have some kind of entanglement dimension).
  Therefore, one might be tempted to declare the dimension of a manifold to be 
the minimal one; it is for now a matter of taste whether one allows for collapse or not and we leave this to the discretion of 
the reader.
\section{Canonical Differentiable Structure}
Before we define a differential structure, we have to identify the natural class of functions on a local chart $(V_{\beta}, \pi_{\beta}, \mathcal{A}_{\beta}, \phi_{\beta})$.  The thing is that points and functions are simply unified in the algebraic context; they just are elements of $\mathcal{A}_{\beta}$.  Indeed, a function is nothing than some limit of a finite polynomial in the $\pi_{\beta}(\widehat{x}_i)$ and the natural question is how we should define the function on an open set $\mathcal{W} \subset \mathcal{V}_{\beta}$.  There are two natural candidates for local functions which we call the entangled and unentangled one for obvious reasons.  The former forgets how an element $A \in \mathcal{A}_{\beta}$ arises from the fundamental building blocks and maps $A \rightarrow \widehat{A}$, where the latter is defined as
$$
\widehat{A}(\mathcal{W}) = P_{\beta}(\mathcal{W})A P_{\beta}(\mathcal{W})\;,
$$
and obviously $\widehat{A}$ maps distinct regions to orthogonal operators; moreover, $\widehat{A}$ preserves the order relation in the sense that
$$
\widehat{\widehat{A}(\mathcal{W}_2)}(\mathcal{W}_1) = \widehat{A}(\mathcal{W}_1)
$$
for $\mathcal{W}_1 \subset \mathcal{W}_2$.  However, this transformation does not erase entanglement with regions outside $\mathcal{W}$ as the reader may easily verify and obviously, this ansatz is not a suitable candidate for defining a differential since it does not ``feel'' the order in which the elementary variables occur.  Let us start with finite polynomials in unity and the preferred variables $\pi_{\beta}(\widehat{x}_i)$, then one meets a rarity which might seem to be a lethal problem at first sight but really is nothing but a manifestation of what breaking of entanglement means.  That is let 
$A = Q(1,\pi_{\beta}(\widehat{x}_i))$, where $Q$ is some polynomial of finite degree, then we define 
$$
\widehat{Q}(\mathcal{W}) = Q(P_{\beta}(\mathcal{W}), P_{\beta}(\mathcal{W}) \pi_{\beta}(\widehat{x}_i) P_{\beta}(\mathcal{W}))
$$
as the local unentangled realization of $Q$.  Now, it is possible for two polynomials $Q_1$ and $Q_2$ to determine identical elements in $\mathcal{A}_{\beta}$, but the local realizations
$\widehat{Q}_j$ differ; also, the reader is invited to construct some examples on this.  All this implies that we have to define nets of polynomials and declare equivalence with respect to the resolution one is measuring which removes all absolutism from $\mathcal{A}_{\beta}$; that is,
$$
\widehat{Q}_1 \sim_{\mathcal{W}} \widehat{Q}_2
$$
if and only if $\widehat{Q}_1(\mathcal{W}) = \widehat{Q}_2(\mathcal{W})$.  One verifies moreover that the local unentangled $\widehat{A}$ has the same inclusion and disjoint properties than the entangled one.  Therefore, consider a natural directed net $(Q_i, i \in \mathbb{N})$ of finite polynomials in the fundamental variables $\widehat{x}_i$ and unity, then we say that the domain $\textrm{Dom}((Q_i, i \in \mathbb{N}), (\pi_{\beta}, \mathcal{A}_{\beta}))$ of this net relative to the chart $(\pi_{\beta}, \mathcal{A}_{\beta})$ is given by the set of relative opens $\mathcal{W} \subset \mathcal{O}(\pi_{\beta}, \mathcal{A}_{\beta})$ so that $\widehat{Q}_i(\mathcal{W})$ is a weakly convergent series of operators.  For the general reader, the weak topology on a $W^{\star}$ algebra is the locally convex topology generated by the continuous complex linear functionals $\psi_{\beta} : \mathcal{A}_{\beta} \rightarrow \mathbb{C}$.  Now in order to define continuity and differentiability of such functions, we need to equip the relative open sets with a canonical topology, that is the Vietoris topology which is defined by the relative open subsets $(\mathcal{O}, \mathcal{V})(\mathcal{W})$ where $\overline{\mathcal{V}} \subset \mathcal{W} \subset \overline{\mathcal{W}} \subset \mathcal{O}$ and $(\mathcal{O}, \mathcal{V})(\mathcal{W})$ is the set of all open sets $\mathcal{Z}$ satisfying $\mathcal{V} \subset \mathcal{Z} \subset \mathcal{O}$.
\newtheorem{theo}{Definition} 
\begin{theo} Therefore, the net $(Q_i, i \in \mathbb{N})$ is of bounded variation relative to $(\pi_{\beta}, \mathcal{A}_{\beta})$ in $\mathcal{W} \in \textrm{Dom}((Q_i, i \in \mathbb{N}), (\pi_{\beta}, \mathcal{A}_{\beta}))$ if and only if for every $\epsilon > 0$ and continuous functional $\psi_{\beta}$, there exists an open set in the Vietoris topology containing $\mathcal{W}$ such that for any open $\mathcal{Z}$ contained in it we have that
$$
|\psi_{\beta}((\widehat{Q}_i, i \in \mathbb{N})(\mathcal{W}) - (\widehat{Q}_i, i \in \mathbb{N})(\mathcal{Z}))| < \epsilon\;.
$$ \end{theo}
In order to define directional continuity, partial differential operators and finite difference operators, we need the notion of directional displacement.  Therefore, let $\vec{e}$ be a unit vector in $\mathbb{R}^n$ and $\delta$; then the translation $T_{(\delta \vec{e})}$ canonically lifts as a continuous map to the space of all open sets by the prescription
$$
T_{(\delta \vec{e})}(\mathcal{W}) = \mathcal{W} + \delta \vec{e}\;.
$$
We need also need to lift the translations to homomorphisms between the local algebras $\mathcal{A}^{loc}_{\beta}(\mathcal{W})$ which requires the use of a quantum connection.  Here, $\mathcal{A}^{loc}_{\beta}(\mathcal{W})$ is the $W^{\star}$ subalgebra of $\mathcal{A}_{\beta}$ generated by $P_{\beta}(\mathcal{W})\pi_{\beta}(\widehat{x}_i)P_{\beta}(\mathcal{W})$ and $P_{\beta}(\mathcal{W})$ which is not the same as $P_{\beta}(\mathcal{W})\mathcal{A}_{\beta} P_{\beta}(\mathcal{W})$ (which is also a Von Neumann algebra) as explained before.  The reason why we need a connection is because at some resolution $\epsilon$, $P_{\beta}(\mathcal{W})$ will not majorize, nor commute with the $P^i((\alpha_i - \epsilon, \alpha_i + \epsilon))$ so that the projection operators will not be projection operators anymore but twisted depending upon the region $\mathcal{W}$ and spectral operator at hand.  This does of course not happen in the abelian case where everything remains trivial.  Also, it is generally not so that for $\mathcal{V} \subset \mathcal{W}$ one obtains that $$\mathcal{A}^{loc}_{\beta}(\mathcal{V}) \subset \mathcal{A}^{loc}_{\beta}(\mathcal{W})$$  and the reason is that fine grained projections can add a twist where coarser grained projections do not.  Of course, this inclusion property does hold when we do not cut entanglement, that is
$$P_{\beta}(\mathcal{V}) \mathcal{A}_{\beta} P_{\beta}(\mathcal{V}) \subset P_{\beta}(\mathcal{W}) \mathcal{A}_{\beta} P_{\beta}(\mathcal{W})$$ for $\mathcal{V} \subset \mathcal{W}$.  Let us give some example confirming these facts, consider the following discrete four dimensional quantum manifold
\begin{eqnarray*}
t & = & \left( \begin{array}{cc}
0 & 1 \\*
1 & 0 \\*	
\end{array} \right) \\
x & = & \left( \begin{array}{cc}
0 & \sigma_1 \\*
\sigma_1 & 0 \\*	
\end{array} \right) \\
y & = & \left( \begin{array}{cc}
0 & \sigma_2 \\*
\sigma_2 & 0 \\*	
\end{array} \right) \\
z & = & \left( \begin{array}{cccc}
2 & 0 & 0 & 0 \\*
0 & 1 & 0 & 0 \\*
0 & 0 & 1 & 0 \\*
0 & 0 & 0 & 0	\\*
\end{array} \right).
\end{eqnarray*} A little algebra reveals that $\left[ t, x \right] = \left[t, y \right] = 0$, $\{ x, y \} = 0$ and 
$t^2 = x^2 = y^2 = 1$; one verifies that these generate $\mathbb{C}^{4 \times 4}$.  One notices that $y$ and $z$ do not commute nor anticommute, the spectrum of $t,x,y$ is $\{ -1, 1 \}$ and both eigenspaces have dimension two; for $z$ it clearly is $\{0,1,2\}$ and therefore the cube consists out of $24$ points.  Associate $\mathcal{V}$ to that subset of the cube with arbitrary values for $t,x$ and $y = 1 = z$ and $\mathcal{W}$ to arbitrary values for $t,x,z$ and $y = 1$, then clearly $\mathcal{V} \subset \mathcal{W}$.  One computes that
\begin{eqnarray*}
P(\mathcal{V}) & = & \frac{1}{2}
\left( \begin{array}{cccc}
0 & 0 & 0 & 0 \\*
0 & 1 & - i & 0 \\*
0 & i & 1 & 0 \\*
0 & 0 & 0 & 0 \\*	
\end{array} \right) 
\end{eqnarray*} 
and $P(\mathcal{W}) = \frac{1}{2} \left( 1 + y  \right)$.  We compute $\mathcal{A}^{loc}(\mathcal{W})$ and show that $P(\mathcal{V})$ does not belong to it.  Elementary algebra shows that
\begin{eqnarray*}
P(\mathcal{W})t P(\mathcal{W}) & = & \frac{1}{2} \left( \begin{array}{cc}
\sigma_2 & 1 \\*
1 & \sigma_2 \\*
\end{array} \right) \\
P(\mathcal{W})x P(\mathcal{W}) & = & 0 \\
P(\mathcal{W})y P(\mathcal{W}) & = & P(\mathcal{W}) \\
P(\mathcal{W})z P(\mathcal{W}) & = & P(\mathcal{W}) \\
\end{eqnarray*} even though $P(\mathcal{W})$ does not commute with $z$.  It is now easy to show that $\mathcal{A}^{loc}(\mathcal{W})$ is two dimensional and that $P(\mathcal{V})$ is not in it.  Finally, we know the dimension of $P(\mathcal{W}) \mathcal{A} P(\mathcal{W})$ is four and the expression
\begin{eqnarray*} \frac{3}{2}z - \frac{1}{2}z^2 & = & \left( \begin{array}{cccc}
1 & 0 & 0 & 0\\*
0 & 1 & 0 & 0 \\*
0 & 0 & 1 & 0 \\*
0 & 0 & 0 & 0 \\*
\end{array} \right) = \alpha \end{eqnarray*}
implies that $P(\mathcal{W}) \alpha P(\mathcal{W}) - \frac{1}{2} P(\mathcal{W}) = \frac{1}{2} P(\mathcal{V})$.    
 \\* \\*
From the weak continuity of $\kappa$ ``almost everywhere'' one deduces that the local algebra's $\mathcal{A}^{loc}(\mathcal{W})$ almost never jump when we move $\mathcal{W}$ around.  Therefore, what one could call quasilocal algebra's are basically the same as the local ones.  Hence, we define a connection, or parallel transport, $\Gamma_{\beta}(\mathcal{V}, \mathcal{W})$ as a bifunction of two relatively open sets which map to a star homomorphism between the respective local algebras; that is,
$$\Gamma_{\beta}(\mathcal{V}, \mathcal{W}) : \mathcal{A}^{loc}_{\beta}(\mathcal{V}) \rightarrow \mathcal{A}^{loc}_{\beta}(\mathcal{W})$$ where a path dependence is possible in the composition and we could at most look for rules of inclusion.  For $\mathcal{V} \subset \mathcal{W}$, one has that when a spectral projector $P \prec P(\mathcal{V})$ or $P(\mathcal{V}) P P(\mathcal{V}) = P$ then the same is true for $P(\mathcal{W})$ and we demand $\Gamma(\mathcal{V}, \mathcal{W})$ to preserve these fixpoints.  Other principles of this kind are not possible, it might be that $P$ commutes with $P(\mathcal{V})$ but not with $P(\mathcal{W})$ and vice versa.  We might still ask however for the connection to be optimal which means that the homomorphisms cannot be majorized.  Therefore, in case the local algebra's are isomorphic, $\Gamma(\mathcal{V}, \mathcal{W})$ is an isomorphism too.  Also, we demand the connection to be unital, meaning that $\Gamma(\mathcal{W}, \mathcal{W})$ is equal to the identity.  There will be two further requirements on the connection which is that the basic functions $\pi_{\beta}(\widehat{x}_i)$ are weakly continuous or differentiable wherever $\kappa_{(\pi_{\beta},\mathcal{A}_{\beta})}$ is in all or some directions $\vec{e}$.  The latter is a huge constraining between the analytical and $W^{\star}$ algebraic aspects of $\mathcal{A}_{\beta}$. \\* \\* 
We have two different notions of continuity and differentiability because $\kappa_{(\pi_{\beta}, \mathcal{A}_{\beta})}$ has a peculiar and natural status within our construction.  First of all, we say that $\kappa_{(\pi_{\beta}, \mathcal{A}_{\beta})}$ is weakly continuous in a point $\mathcal{W}$ in the Vietoris topology when for all $\epsilon > 0$ and continuous functionals $\psi_{\beta}$, there exists an open neighborhood $\mathcal{O}$ in the Vietoris topology such that for any $\mathcal{Z} \in \mathcal{O}$ we have that
$$| \psi_{\beta} \left( \kappa_{(\pi_{\beta}, \mathcal{A}_{\beta})}(\mathcal{W}) - \kappa_{(\pi_{\beta}, \mathcal{A}_{\beta})}(\mathcal{Z}) \right) | < \epsilon.$$  Likewise, we say that $\kappa_{(\pi_{\beta}, \mathcal{A}_{\beta})}$ is continuous in the direction $\vec{e}$ at $\mathcal{W}$ when for any $\epsilon > 0$ and $\psi_{\beta}$, there exists a $\delta > 0$ so that for any $|h| < \delta$ 
$$|\psi_{\beta} \left( \kappa_{(\pi_{\beta}, \mathcal{A}_{\beta})}(\mathcal{W}) - \kappa_{(\pi_{\beta}, \mathcal{A}_{\beta})}(T_{(h\vec{e})}(\mathcal{W})) \right) | < \epsilon.$$
Concerning the notion of weak differentiability of $\kappa_{(\pi_{\beta}, \mathcal{A}_{\beta})}$, there exist several and we have to find out if some of them are equivalent or not.  Let me first start by examining the abelian case in sufficient detail and then generalize to the nonabelian setting.  In the Schrodinger like setting explained before, the projection operators are just characteristic functions and in one dimension, the computations simplify considerably (however, there is no problem generalizing this to higher dimensions as the reader may try to do) while the results are universal.  Naively, one would think we have to calculate the limit of
$$\frac{1}{\delta} \left( \chi_{(a + \delta, b + \delta)} - \chi_{(a,b)} \right)$$ for $0 < \delta \rightarrow 0$.  If one would restrict to the continuous functions as a separating \emph{subalgebra} of the $L^{2}$ functions (at least on a compact measure space), then this limit exists in the weak sense and it is $\delta(b) - \delta(a)$ which is outside the algebra since it is not well defined on the whole Hilbert space.  Now, if again, we would only restrict to the continuous functions, then the limit  
$$\frac{1}{\delta^{1 - \gamma}} \left( \chi_{(a + \delta, b + \delta)} - \chi_{(a,b)} \right)$$ is zero and independent of $\gamma > 0$.  However, if one were to go over to the full Hilbert space, then it is necessary and sufficient that $\gamma > \frac{1}{2}$ in which case the limit is also zero.  Therefore, we say that $\kappa_{(\pi_{\beta}, \mathcal{A}_{\beta})}$ is $\gamma$-weakly differentiable with respect to a separating\footnote{Separating means that for all distinct $A,B \in \mathcal{A}_{\beta}$ there exists a $\psi_{\beta} \in \Psi_{\beta}(\gamma)$ such that $\psi_{\beta}(A) \neq \psi_{\beta}(B)$.} subset $\Psi_{\beta}(\gamma)$ of continuous functionals in the direction $\vec{e}$ at $\mathcal{W}$ if there exists an element $\partial^{\gamma}_{\vec{e}} \, \kappa_{(\pi_{\beta}, \mathcal{A}_{\beta})}(\mathcal{W})$  such that for all $\epsilon > 0$ and $\psi_{\beta} \in \Psi_{\beta}(\gamma)$, there exists a $\delta > 0$ such that for all $0 < h < \delta$ we have that 
$$\left|\psi_{\beta} \left( \frac{1}{h^{1 - \gamma}} \left( \kappa_{(\pi_{\beta}, \mathcal{A}_{\beta})}(T_{(h\vec{e})}(\mathcal{W})) - \kappa_{(\pi_{\beta}, \mathcal{A}_{\beta})}(\mathcal{W}) \right) - \partial^{\gamma}_{\vec{e}} \, \kappa_{(\pi_{\beta}, \mathcal{A}_{\beta})}(\mathcal{W}) \right)  \right| <  \epsilon.$$
Similarly, one could forget about $\Psi_{\beta}(\gamma)$ and demand that $\gamma > \frac{1}{2}$.  This attitude could lead to very different algebra's and we will not even start its investigation in this short paper.  An obvious property is that if $\kappa_{(\pi_{\beta}, \mathcal{A}_{\beta})}$ is differentiable with respect to $(\gamma_1, \Psi_{\beta}(\gamma_1))$, then it is also the case for $(\gamma_2, \Psi_{\beta}(\gamma_1))$ where $\gamma_2 < \gamma_1$ and the differential is exactly zero. \\* \\*
We now turn to continuity and differentiability of nets $(\widehat{Q}_i , i \in \mathbb{N})$ of finite polynomials on their relative domain (with respect to $(\pi_{\beta}, \mathcal{A}_{\beta})$).  Define now $$\widehat{T}_{(\delta \vec{e})}(\mathcal{W}) = \Gamma(\mathcal{W}, T_{(\delta \vec{e})}(\mathcal{W}))$$ then we say that $(\widehat{Q}_i , i \in \mathbb{N})$ differentiable at $\mathcal{W}$ in the interior of its relative domain in the direction of $\vec{e}$ if and only if there exists a unique element $$\partial_{\vec{e}} \,(\widehat{Q}_i, i \in \mathbb{N})(\mathcal{W}) \in \mathcal{A}^{loc}(\mathcal{W})$$ such that for any $\psi_{\beta}$,
\begin{eqnarray}
& &\psi_{\beta} \left( \partial_{\vec{e}} \, (\widehat{Q}_i, i \in \mathbb{N})(\mathcal{W}) \right) \nonumber\\
& &=\ \lim_{\delta \rightarrow 0}  \frac{1}{\delta} \, \psi_{\beta} \left( \widehat{T}_{- \delta \vec{e}} \left[(\widehat{Q}_i, i \in \mathbb{N})(T_{(\delta \vec{e})}(\mathcal{W})) \right] - (\widehat{Q}_i, i \in \mathbb{N})(\mathcal{W}) \right). \nonumber
\end{eqnarray}
So, the differential operator is only defined if some translates of $\mathcal{W}$ belong to the relative domain of $(\widehat{Q}_i, i \in \mathbb{N})$ for arbitrarily small $\delta$.  Therefore, partial differentials are not defined for directions in which the set at hand is isolated.  Of course, if one looks only at larger scales, then jumps may be accomplished and the difference operators are canonically defined.  One could also resort here to notions of $(\gamma, \Psi_{\beta}(\gamma))$ differentiability, but I see no stringent need to do it at this point. \\* \\* 
Before we give some examples, let us proceed by defining the holonomy groups attached to the connection;  for any $\mathcal{W}$, we define $H(\mathcal{W})$ as the group of homomorphisms from $\mathcal{A}^{loc}(\mathcal{W})$ to itself generated by finite compositions of the kind $$\Gamma(\mathcal{W}_n, \mathcal{W}) \Gamma(\mathcal{W}_{n-1}, \mathcal{W}_n) \ldots \Gamma(\mathcal{W}_{1}, \mathcal{W}_{2})\Gamma(\mathcal{W}, \mathcal{W}_1).$$
We say that a connection is flat when all the holonomy groups are equal to the identity.  Consider as before the trivial example of a real $n$ dimensional manifold, then the translation mappings induce a canonical flat connection on the pairs of opens differing by a translate as follows: every spectral operator $P((\alpha_i - \epsilon, \alpha_i + \epsilon) \cap \mathcal{W}) = P(\mathcal{W}) P^{i}((\alpha_i - \epsilon, \alpha_i + \epsilon))P(\mathcal{W})$ gets mapped to $$P((\alpha_i - \epsilon + \delta e_i, \alpha_i + \epsilon + \delta e_i) \cap T_{(\delta \vec{e})}(\mathcal{W}))$$ provided $\mathcal{W}$ and $\mathcal{W} + \delta \vec{e}$ belong to the cube.  Actually, this is all we need to calculate differentials and so on, but the reader might wish to extend this definition in a canonical way to generic pairs.  For $\mathcal{W}$ of compact closure and real differentiable function $f$ (with uniformly continuous partial derivatives) with $\overline{\mathcal{W}} \subset \textrm{Dom}(f)$ one associates a unique algebra element $\widetilde{f}$ (in the commutative case we do not need the nets).  It is easy to calculate that the new differential
$$
\partial_{\vec{e}} \, \widehat{\widetilde{f}}(\mathcal{W}) = \widetilde{\partial_{\vec{e}} \, f \, \chi_{\mathcal{W}}}\;,
$$
reduces to the old one and that the latter even exists in the norm topology in this case.  \\* \\*
All these results allow us now to obtain a better insight into the nature of noncommutative $n$ dimensional manifolds.  Before we engage in this discussion we still need to solve some questions:  
\begin{itemize} 
\item We have demanded that for overlapping charts the algebra's of local projection operators (with respect to these charts) are isomorphic; how does this algebra relate to the local algebra with respect to that chart?
\item We have seen that for $\mathcal{V} \subset \mathcal{W}$, it does not necessarily hold that $\mathcal{A}^{loc}(\mathcal{V}) \subset \mathcal{A}^{loc}(\mathcal{W})$.  However, does there exist an isomorphism of $\mathcal{A}^{loc}(\mathcal{V})$ into a subalgebra of $\mathcal{A}^{loc}(\mathcal{W})$ ?
\item  Finally, say that $\mathcal{W}$ contains $r$ components with respect to $\mathcal{V}_{\beta}$; does the spectrum of the local algebra $\mathcal{A}^{loc}(\mathcal{W})$ contain at least $r$ components ?  
\end{itemize} 
As a response to the first question, we already know that the algebra of local projection operators is not contained in the local algebra and the question is whether the inverse holds.  But before we treat these questions in generality, let us see how they are answered in the our previous example.  Concerning the first question, we notice that the only nonzero projection operators (apart from $P(\mathcal{V})$ and $P(\mathcal{W})$) arise from $y = 1$ and $t = \pm 1$; they are given by
\begin{eqnarray*}
P(t = 1 = y) & = & \frac{1}{4}
\left( \begin{array}{cccc}
1 & i & 1 & i \\*
- i & 1 & - i & 1 \\*
1 & i & 1 & i \\*
-i & 1 & -i & 1	
\end{array} \right) \\
P(t = -1 = -y) & = & \frac{1}{4} \left(
\begin{array}{cccc}
1 & -i & -1 & i \\*
i & 1 & - i & -1 \\*
-1 & i & 1 & -i \\*
-i & -1 & i & 1	
\end{array} \right).
\end{eqnarray*}
It is most easily seen that $P(\mathcal{W})tP(\mathcal{W}) = 2 P(t = 1 = y) - P(\mathcal{W})$ which shows that $\mathcal{A}^{loc}(\mathcal{W})$ is a subalgebra of the algebra generated by the local projection operators $P(\mathcal{V})$ with $\mathcal{V} \subseteq \mathcal{W}$.  The second question is answered in the \emph{negative}  since $\mathcal{A}^{loc}(\mathcal{V})$ is generated by $P(\mathcal{V})$ and
\begin{eqnarray*}
P(\mathcal{V}) t P(\mathcal{V}) & = & \frac{1}{2} \left(
\begin{array}{cccc}
1 & 0 & 0 & i \\*
0 & 0 & 0 & 0 \\*
0 & 0 & 0 & 0 \\*
- i & 0 & 0 & 1	
\end{array} \right) 
\end{eqnarray*} and it is easy to verify that this algebra is not isomorphic to $\mathcal{A}^{loc}(\mathcal{W})$.
Therefore, the answer to the second question is inconclusive since in the commutative case $\mathcal{A}^{loc}(\mathcal{V}) \subseteq \mathcal{A}^{loc}(\mathcal{W})$.  Regarding the third issue, $\mathcal{W}$ contains $12$ points and the cube of $\mathcal{A}^{loc}(\mathcal{W})$ contains also $12$ of them\footnote{One calculates that the spectrum of $P(\mathcal{W})tP(\mathcal{W})$ is $\{ - 2, 2, 0\}$ and the projection operator on the zero eigenvalue is $\frac{1}{2}( 1 - y)$.}.  However, all projection operators vanish in the former case while in the latter exactly $3$ of them are nonzero.  Therefore, the question appears to hold on the ontological as well as the empirical level. \\* \\*  Let us start with some mathematical preliminaries.
\newtheorem{theor}{Theorem} 
\begin{theor} Let $P$ and $Q$ be two (noncommuting) Hermitian projection operators then the projection operators $P \wedge Q$ and $P \vee Q$ belong to $\mathcal{M}' \cap \mathcal{M}$, where $\mathcal{M}'$ is the commutant in $\mathcal{A}_{\beta}$ of the Von Neumann algebra $\mathcal{M}$ generated by $P$ and $Q$.  In particular, any Hermitian projection operator which is smaller than $P \wedge Q$ or larger than $P \vee Q$ belongs to $\mathcal{M}'$.
\end{theor} 
\textit{Proof} : Represent $P$ and $Q$ on a Hilbert space $\mathcal{H}$ and consider the smallest closed subspace $\mathcal{H}'$ which is left invariant by both of them.  Then this $\mathcal{H}'$ has $P \vee Q$ as identity operator and we have to show that it is generated by $P$ and $Q$.
For the intersection, the proof is easy: $\frac{1}{2} \left( PQ + QP \right) = P \wedge Q + A$ where $(P \wedge Q)A = 0$, $A^{\star} = A$, $|| A || \leq 1$ but $1$ does not belong to the discrete spectrum, and therefore
$$P \wedge Q = \lim_{n \rightarrow \infty} \left( \frac{1}{2} (PQ + QP) \right)^{n}$$ in the weak sense.  Replacing $Q$ by $Q' = Q - PQ - QP + PQP$, we see that it is zero if and only if $Q = P$; moreover, $PQ' = Q'P = 0$ and $Q'$ as a mapping from $(1 - P)\mathcal{H}'$ to $(1 - P)\mathcal{H}'$ does not contain $0$ in its discrete spectrum.  Otherwise, there would exist a vector $v \in (1-P)\mathcal{H}'$ such that $(1 - P)Qv = 0$ or $Qv = PQv$ which is impossible unless $v$ is in the intersection of both hyperspaces which implies it must be the zero vector.  In the finite dimensional case, it easy to construct polynomials $f_{\alpha}(x)$ with $f_{\alpha}(0) = 0$ such that $$f_{\alpha}(Q') = P_{\alpha}$$ where $\alpha \in \sigma(Q')$ and $P_{\alpha}$ is its spectral operator.  Therefore, one can recuperate the identity $P \vee Q - P$ on $(1 - P)\mathcal{H}'$ in the algebra of $Q'$ only.  In the infinite dimensional case, this technique fails since the polynomials will start to oscillate heavily which has a detrimental effect on the continuous spectrum.   However, if one considers the algebra generated by $1,P,Q$ a similar argument holds due to the Stone Weierstrass and spectral theorem. \\* \\*
Concerning the first question, let us elaborate on whether given a cube $P_1, P_2$ where $P_1 = P + Q$ with $PQ = 0$ and corresponding to distinct discrete eigenvalues, it is true that
$$(P_1 \wedge P_2) P (P_1 \wedge P_2) = \alpha P_1 \wedge P_2 + (1 - \alpha)P \wedge P_2 - \alpha Q \wedge P_2$$
for some $\alpha \in \mathbb{R}$ (actually the reader can check that any linear combination of these operators has to be of this form).  It is easily seen that this statement is false, since consider the orthonormal unit vectors $e_i$, $i : 1 \ldots 5$, and the following subspaces:
\begin{eqnarray*}
\mathcal{P} & = & \textrm{Span} \{ \cos (\theta) e_1 + \sin (\theta) e_2, \cos (\psi)e_3 + \sin (\psi) e_4  \} \\*
\mathcal{Q} & = & \textrm{Span} \{ \sin (\theta)e_1 - \cos (\theta) e_2 , \sin (\psi)e_3 - \cos(\psi) e_4 \} \\*
\mathcal{P}_2 & = & \textrm{Span} \{ e_2, e_3 , e_5 \}.
\end{eqnarray*} Then, one has the following identities: 
\begin{eqnarray*}
P_1 \wedge P_2 & = & | e_2 \rangle \langle e_2 | + | e_3 \rangle \langle e_3 | \\*
P \wedge P_2 & = & 0 \\*
Q \wedge P_2 & = & 0 \\*
PQ & = & 0.
\end{eqnarray*} However, one easily calculates that
\begin{eqnarray*}
(P_1 \wedge P_2) P (P_1 \wedge P_2)  & = & \sin^2(\theta) |e_2 \rangle \langle e_2 | + \cos^2(\psi) | e_3 \rangle \langle e_3 |
\end{eqnarray*} which is not a multiple of $P_1 \wedge P_2$.  Therefore, one has that $P(\mathcal{W}) P^i P(\mathcal{W})$ is in general not in the algebra generated by $P(\mathcal{V})$ where $\mathcal{V} \subseteq \mathcal{W}$.  It is now easy to pick $\pi_{\beta}(\widehat{x}_i) = P + \mu R$ where $R = |e_5 \rangle \langle e_5|$ to conclude that $$(P_1 \wedge P_2) \pi_{\beta}(\widehat{x}_i) (P_1 \wedge P_2)$$ is not in the algebra generated by the $P(\mathcal{V})$.  This shows that $\mathcal{A}^{loc}(\mathcal{W})$ and the $W^{\star}$ algebra  $\mathcal{A}^{open}(\mathcal{W})$ generated by the $P(\mathcal{V})$ where $\mathcal{V} \subseteq \mathcal{W}$ have no relation to one and another.
\begin{theo} We call the chart $(\mathcal{V}_{\beta}, \pi_{\beta}, \mathcal{A}_{\beta} , \phi_{\beta})$ \emph{pointed} when for all $\mathcal{W}$, $$\mathcal{A}^{loc}(\mathcal{W}) \subseteq \mathcal{A}^{open}(\mathcal{W}).$$
\end{theo}  
We now proceed to answer the third question which intuitively means that if you zoom in you see more and more disconnected components.  Now, it is obvious that this property does not even hold in the commutative case where on large scales one may see many isles but on small scales all one sees is one of them.  However, a refinement of the question is nevertheless interesting and one might want to look for manifolds which have only one component on a given scale and where the number of components grows polynomially (or even exponentially) in the inverse scaling $\frac{1}{\lambda}$.  \\*  \\*
We now have obtained a better view on how we should do function theory on a noncommutative topological manifold although we are confronted with an apparent dilemma.  On one side $\mathcal{A}^{open}_{\beta}(\mathcal{W})$ is the natural algebra we should use to compare overlapping charts, but $\mathcal{A}^{loc}_{\beta}(\mathcal{W})$ is the natural algebra for function theory.  What we learned is that they have generically little to do with one and another; therefore, this begs the question of how to even define \emph{algebraic} functions on the entire manifold.  It is here that the (trace) functionals $\omega_{\mathcal{A}_{\beta}}$ come into play in the following sense: let $\mathcal{M}$ be a noncommutative manifold, then $F: \tau(\mathcal{M}) \rightarrow \mathbb{C}$, where $\tau(\mathcal{M})$ is the set of open subsets of $\mathcal{M}$ equipped with the Vietoris topology, is an algebraic function if and only if for any chart $(\mathcal{V}_{\beta}, \pi_{\beta}, \mathcal{A}_{\beta} , \phi_{\beta})$, there exists a net of polynomials $(Q^{\beta}_i , i \in \mathbb{N})$ such that $$F(\mathcal{W}) = \omega_{\mathcal{A}_{\beta}} \left(  \left( \widehat{Q}^{\beta}_{i} , i \in \mathbb{N} \right)(\mathcal{W}) \right).$$  Continuity of $F$ is obviously defined with respect to the Vietoris topology.  We call $F$ nuclear if and only if for any $\mathcal{V}, \mathcal{W}$, one has that 
$$F(\mathcal{V} \cup \mathcal{W}) = F(\mathcal{W}) + F(\mathcal{V}) - F(\mathcal{V} \cap \mathcal{W}).$$  
Obviously, the standard continuous functions on a real $n$ dimensional manifold with a volume element induce nuclear continuous functions
 by putting the trace functional equal to the $n$ dimensional integral.  We can define higher order algebraic functions as follows 
$$F(\mathcal{W}, \mathcal{V}_1, \ldots \mathcal{V}_m) = \omega_{\mathcal{A}_{\beta}} \left( P_{\beta}(\mathcal{V}_1 ) \ldots P_{\beta}(\mathcal{V}_m) \left( \widehat{Q}^{\beta}_i , i \in \mathbb{N} \right)(\mathcal{W}) \right)$$ where
$\mathcal{V}_j  \subset \mathcal{W}$.  
The gluing conditions ensure us that the identity element in $\mathcal{F}_n$ canonically defines a set of (higher order) algebraic 
functions.  One could now study trace abelian local representations of algebraic functions $F$; more specifically, consider any two overlapping
charts $(\mathcal{V}_{\beta_i},\pi_{\beta_i}, \mathcal{A}_{\beta_i},\phi_{\beta_i})$ and consider any open $\mathcal{W} \subset 
\mathcal{V}_{\beta_1} \cap \mathcal{V}_{\beta_2}$.  We demand there to exist a matrix valued function $h^{k}_{l}(\beta_{1},\beta_{2})$
 on the topology such that 
$$\omega_{\beta_1} \left( \widehat{\partial}_{k} \left( \widehat{Q}^{1}_{j} ; j \in \mathbb{N} \right)(\mathcal{W}) \right) = 
\omega_{\beta_2}  \left( \int_{\phi_{\beta_1}(\mathcal{W})} \frac{\partial^l (\phi_{\beta_2} \circ \phi_{\beta_1}^{-1})(x)}{ \partial x^k}
h^{r}_{l}(\beta_1,\beta_2)(x) \widehat{\partial}_{r} \left( \widehat{Q}^{2}_{j}; j \in \mathbb{N} \right)(x) \right) $$
where the integral is understood to be taken in some ordered sense by evaluating the functions on (almost everywhere) partitions by open subsets.  For 
standard abelian manifolds and nuclear functions, the standard matrix $h^{k}_{l}(\beta_1,\beta_2)$ is given by
$$h^{k}_{l}(\beta_1,\beta_2)(\mathcal{W}) = \delta^{k}_{l} \frac{1}{\textrm{Vol}_{\beta_2}(\mathcal{W})}.$$   
Let us finish by commenting upon the very act of pasting together ``algebraic charts''. 
  We have learned two ways of cutting entanglement, which was by going over to local and open $W^{\star}$ algebra's associated to open 
subsets of $\mathcal{M}$; also, the $W^{\star}$ algebraic framework forces us in the cauldron of relatively open subsets of
 $\mathbb{R}^n$.  This implies that in order to generate a nontrivial topology (with respect to a continuum background) some sort of 
``decoherence'' has to occur.  Indeed, saying that two charts are described by separate $W^{\star}$ algebra's really means that the 
points in both charts do not ``entangle'' in some sense.  Whether or not this is a desirable conclusion remains to be seen. 
\section{Conclusions}
We have made first steps with universal n dimensional manifolds in the context
of $W^{\star}$ algebra's by defining topological noncommutative manifolds, clarifying
the (lack of) relationships between different local $W^{\star}$ algebra's and by mak-
ing first steps with functional calculus. What remains to be done it to treat
higher differential calculus and define general differentiable nonabelian manifolds. From thereon, one can construct vector and tensor calculus and define
noncommutative geometry. It would be instructive to construct explicit realizations of Hopf algebra's as our construction should allow for this and much
more; this would offer a concrete interpretational framework for amongst others
kappa Minkowski spacetime.
Hence, the we have reached the conclusion that what we see depends upon the
scale that we are looking at, but the continuum $\mathbb{R}^n$ background always is and
constitutes the very backbone of the entire construction. Therefore, spacetime
is grounded in the continuum albeit we may perceive it in an atomistic way.
This is precisely the conclusion the author advocated in \cite{Noldus2}.
\section{Acknowledgements}
J. Noldus expresses his gratitude to Steve Adler for the nice and productive in-
vitation at the IAS at Princeton University. Also people at the math institute
in Utrecht have provided valuable comments including Jan Sienstra
and Andr\'e Henriques.  Finally, he is indebted to Wilhem Furtwangler and Kirsten Flagstad for inspiration. The very existence of
this paper is due to some ``bet'' with Daniele Oriti about ``how a chicken should
be cooked'' and he is acknowledged for giving me the opportunity to write down
the recipe.


\begin{thebibliography}{99}
\bibitem{Noldus1} J. Noldus, Foundations of a theory of quantum gravity, Vixra:1106.0028 
\bibitem{Noldus2} J. Noldus, How to learn to ask good questions in physics?, paper for FQXi
    contest, Vixra:1108.0026 
\bibitem{Majid1} S. Majid, A Quantum Groups Primer, London Mathematical Society Lec-
    ture Notes, vol 292, Cambridge University Press (2002) 182pp 
\bibitem{Majid2} S. Majid, Foundations of Quantum Group Theory, Cambridge University
    Press (1995) 1-640pp (2nd edn 2000) 
\bibitem{Kleinert} Demazure Michel and Alexandre Grothendieck : S\'eminaire de G\'eometrie
    Alg\'ebrique du Bois Marie - 1962-64 - Sch\'emas en groupes - (SGA 3) - (Lecture notes in mathematics 151). Berlin; New York: Springer-Verlag.
pp. xv+564.
\end{thebibliography}
\end{document}